\documentclass[12pt]{article}
\usepackage{amssymb}
\usepackage{graphicx}
\begin{document}
\newcommand{\beq}{\begin{equation}}
\newcommand{\eeq}{\end{equation}}
\newcommand{\beqa}{\begin{eqnarray}}
\newcommand{\eeqa}{\end{eqnarray}}
\newcommand{\beqar}{\begin{eqnarray*}}
\newcommand{\eeqar}{\end{eqnarray*}}
\newcommand{\al}{\alpha}
\newcommand{\be}{\beta}
\newcommand{\del}{\delta}
\newcommand{\D}{\Delta}
\newcommand{\eps}{\epsilon}
\newcommand{\ga}{\gamma}
\newcommand{\Ga}{\Gamma}
\newcommand{\ka}{\kappa}
\newcommand{\nn}{\nonumber}
\newcommand{\inn}{\!\cdot\!}
\newcommand{\h}{\eta}
\newcommand{\ii}{\iota}
\newcommand{\kk}{\varphi}
\newcommand\F{{}_3F_2}
\newcommand{\la}{\lambda}
\newcommand{\La}{\Lambda}
\newcommand{\na}{\prt}
\newcommand{\Om}{\Omega}
\newcommand{\om}{\omega}
\newcommand{\p}{\Phi}
\newcommand{\sig}{\sigma}
\renewcommand{\t}{\theta}
\newcommand{\z}{\zeta}
\newcommand{\ssc}{\scriptscriptstyle}
\newcommand{\eg}{{\it e.g.,}\ }
\newcommand{\ie}{{\it i.e.,}\ }
\newcommand{\labell}[1]{\label{#1}} 
\newcommand{\reef}[1]{(\ref{#1})}
\newcommand\prt{\partial}
\newcommand\veps{\varepsilon}
\newcommand{\pol}{\varepsilon}
\newcommand\vp{\varphi}
\newcommand\ls{\ell_s}
\newcommand\cF{{\cal F}}
\newcommand\cA{{\cal A}}
\newcommand\cS{{\cal S}}
\newcommand\cT{{\cal T}}
\newcommand\cV{{\cal V}}
\newcommand\cL{{\cal L}}
\newcommand\cM{{\cal M}}
\newcommand\cN{{\cal N}}
\newcommand\cG{{\cal G}}
\newcommand\cK{{\cal K}}
\newcommand\cH{{\cal H}}
\newcommand\cI{{\cal I}}
\newcommand\cJ{{\cal J}}
\newcommand\cl{{\iota}}
\newcommand\cP{{\cal P}}
\newcommand\cQ{{\cal Q}}
\newcommand\cg{{\it g}}
\newcommand\cR{{\cal R}}
\newcommand\cB{{\cal B}}
\newcommand\cO{{\cal O}}
\newcommand\tcO{{\tilde {{\cal O}}}}
\newcommand\bz{\bar{z}}
\newcommand\bb{\bar{b}}
\newcommand\ba{\bar{a}}
\newcommand\bg{\bar{g}}
\newcommand\bc{\bar{c}}
\newcommand\bomega{\bar{\omega}}
\newcommand\bH{\bar{H}}
\newcommand\bw{\bar{w}}
\newcommand\bX{\bar{X}}
\newcommand\bK{\bar{K}}
\newcommand\bA{\bar{A}}
\newcommand\bR{\bar{R}}
\newcommand\bZ{\bar{Z}}
\newcommand\bxi{\bar{\xi}}
\newcommand\bphi{\bar{\phi}}
\newcommand\bpsi{\bar{\psi}}
\newcommand\bprt{\bar{\prt}}
\newcommand\bet{\bar{\eta}}
\newcommand\btau{\bar{\tau}}
\newcommand\hF{\hat{F}}
\newcommand\hA{\hat{A}}
\newcommand\hT{\hat{T}}
\newcommand\htau{\hat{\tau}}
\newcommand\hD{\hat{D}}
\newcommand\hf{\hat{f}}
\newcommand\hK{\hat{K}}
\newcommand\hg{\hat{g}}
\newcommand\hp{\hat{\Phi}}
\newcommand\hi{\hat{i}}
\newcommand\ha{\hat{a}}
\newcommand\hb{\hat{b}}
\newcommand\hQ{\hat{Q}}
\newcommand\hP{\hat{\Phi}}
\newcommand\hS{\hat{S}}
\newcommand\hX{\hat{X}}
\newcommand\tL{\tilde{\cal L}}
\newcommand\hL{\hat{\cal L}}
\newcommand\MZ{\mathbb{Z}}
\newcommand\MR{\mathbb{R}}
\newcommand\tG{{\tilde G}}
\newcommand\tg{{\tilde g}}
\newcommand\tphi{{\widetilde \Phi}}
\newcommand\tPhi{{\widetilde \Phi}}
\newcommand\ti{{\tilde i}}
\newcommand\tj{{\tilde j}}
\newcommand\tk{{\tilde k}}
\newcommand\tl{{\tilde l}}
\newcommand\ttm{{\tilde m}}
\newcommand\tn{{\tilde n}}
\newcommand\ta{{\tilde a}}
\newcommand\tb{{\tilde b}}
\newcommand\tc{{\tilde c}}
\newcommand\td{{\tilde d}}
\newcommand\tm{{\tilde m}}
\newcommand\tmu{{\tilde \mu}}
\newcommand\tnu{{\tilde \nu}}
\newcommand\talpha{{\tilde \alpha}}
\newcommand\tbeta{{\tilde \beta}}
\newcommand\trho{{\tilde \rho}}
 \newcommand\tR{{\tilde R}}
\newcommand\teta{{\tilde \eta}}
\newcommand\tF{{\widetilde F}}
\newcommand\tK{{\widetilde K}}
\newcommand\tE{{\tilde E}}
\newcommand\tpsi{{\tilde \psi}}
\newcommand\tX{{\widetilde X}}
\newcommand\tD{{\widetilde D}}
\newcommand\tO{{\widetilde O}}
\newcommand\tS{{\tilde S}}
\newcommand\tB{{\tilde B}}
\newcommand\tA{{\widetilde A}}
\newcommand\tT{{\widetilde T}}
\newcommand\tC{{\widetilde C}}
\newcommand\tV{{\widetilde V}}
\newcommand\thF{{\widetilde {\hat {F}}}}
\newcommand\Tr{{\rm Tr}}
\newcommand\tr{{\rm tr}}
\newcommand\STr{{\rm STr}}
\newcommand\hR{\hat{R}}
\newcommand\M[2]{M^{#1}{}_{#2}}
\newcommand\dB{\dot{B}}
\newcommand\dG{\dot{G}}
\newcommand\ddG{\dot{\dot{G}}}
\newcommand\dP{\dot{\phi}}
\newcommand\ddB{\dot{\dot{B}}}
\newcommand\ddp{\dot{\dot{\Phi}}}

\newcommand\bS{\textbf{ S}}
\newcommand\bI{\textbf{ I}}
\newcommand\bJ{\textbf{ J}}
\newcommand\bL{\textbf{\cL}}

\begin{titlepage}
\begin{center}

\vskip 2 cm
{\LARGE \bf  T-duality constraint on effective Lagrangians  }\\
\vskip 1.25 cm
   Mohammad R. Garousi\footnote{garousi@um.ac.ir}

\vskip 1 cm
{{\it Department of Physics, Faculty of Science, Ferdowsi University of Mashhad\\}{\it P.O. Box 1436, Mashhad, Iran}\\}
\vskip .1 cm
 \end{center}

\begin{abstract}

  Recent studies have highlighted the significant role of utilizing $O(1,1)$ symmetry in the circular reduction of effective actions to determine NS-NS couplings in the effective action of string theory. However, these calculations often result in residual terms as total derivatives that do not conform to $O(1,1)$ transformations. In this paper, we present explicit calculations at $\alpha'$ order, demonstrating the enforceability of this symmetry on effective Lagrangians to establish the parameters governing covariant couplings in any scheme. Notably, we discover the $O(1,1)$-invariant Lagrangians corresponding to the Metsaev-Tseytlin action and the Meissner action.

\end{abstract}
\end{titlepage}

\section{Introduction}

It is well-known that the classical effective action of string theory, when dimensionally reduced on a torus $T^{(d)}$, exhibits a global symmetry called $O(d,d)$ \cite{Sen:1991zi, Hohm:2014sxa}. By assuming that the effective action of string theory at the critical dimension is background independent, one can consider a background with toroidal compactification $T^{(d)}$ and impose the non-geometric subgroup of the $O(d,d)$ symmetry on the reduction of the most general covariant and gauge-invariant couplings. Each coupling has its own arbitrary coefficients, and this allows for determining the connections between the coefficients of the independent couplings.

This approach has been applied in previous works \cite{Garousi:2019mca, Garousi:2023kxw, Garousi:2020gio} for the case of $d=1$, where connections between the NS-NS couplings at order $\alpha'^2$ in bosonic and heterotic theories, as well as the couplings at order $\alpha'^3$ in superstring theory, were established. In these calculations, the initial steps involve employing the Bianchi identities, the most general higher-derivative field redefinitions, and eliminating total derivative terms from the most general covariant action to obtain the minimal independent couplings in the action in a particular scheme \cite{Metsaev:1987zx}. The subsequent application of the non-geometric part of the $O(1,1)$ symmetry, known as the Buscher rules \cite{Buscher:1987sk, Buscher:1987qj}, determines the independent couplings up to only one unambiguous parameter, which can be fixed by other means such as the S-matrix method. Furthermore, these calculations extended the Buscher rules to incorporate higher-derivative corrections that depend on the scheme of the covariant couplings \cite{Garousi:2019wgz}. However, the circular reduction of the couplings obtained in \cite{Garousi:2019mca, Garousi:2023kxw, Garousi:2020gio} is $O(1,1)$ invariant, except for some anomalous terms in the form of total derivative terms in the base space.
In closed spacetime manifolds, the anomalous total derivative terms are negligible, leading to effective actions that exhibit invariance under $O(1,1)$ transformations. The background independence assumption of the effective action suggests that this symmetry should also hold in open spacetime manifolds. However, in these manifolds, the total derivative terms cannot be ignored. Therefore, one can approach the problem in two ways.

The first approach involves utilizing Stokes' theorem to transfer the anomalous total derivative terms to the boundary. By introducing appropriate boundary terms that undergo $O(1,1)$ transformations and possess anomalies under these transformations, it is possible to cancel out the aforementioned anomalous terms on the boundary. In this way, one may be able to find the Gibbons-Hawking-like boundary terms \cite{Gibbons:1976ue,Garousi:2019xlf,Garousi:2021cfc,Garousi:2021yyd}.

Alternatively, there is a Lagrangian corresponding to each effective action. This Lagrangian may maintain invariance under $O(1,1)$ transformations without any anomaly. This paper focuses on exploring this latter possibility, highlighting the existence of Lagrangians for the effective actions in each scheme that satisfy $O(1,1)$ symmetry.

The most general covariant and gauge-invariant Lagrangians in the bulk, at order $\alpha'^m$, can contain terms with up to $2m+2$ derivatives. The coefficients of these couplings are interconnected only through the Bianchi identities and field redefinitions. By exploiting the Bianchi identities, one can reduce the couplings that are solely related to each other by these identities to a set of independent couplings. However, the coefficients of the resulting couplings are either unambiguous, meaning they are invariant under field redefinitions, or they are ambiguous, meaning they change under field redefinitions. Moreover, under field redefinitions, the ambiguous parameters are further divided into essential parameters and arbitrary parameters. The number of essential parameters at each order of $\alpha'$ is fixed. The choice of which set of ambiguous parameters is selected as essential parameters specifies the scheme of the Lagrangian. If one sets all arbitrary parameters to zero, the unambiguous and essential couplings in the Lagrangian then appear in the minimal scheme. The other choices for the arbitrary parameters correspond to other schemes that are related to the minimal scheme by field redefinitions. Note that the number of couplings in the Lagrangian in the minimal scheme is greater than the number of couplings in the action in the minimal scheme  \cite{Metsaev:1987zx}, in which the total derivative terms are also removed by using integration by parts.

In the case where field redefinitions are used, the $O(1,1)$ symmetry can determine all parameters in the Lagrangian in the minimal scheme and all parameters in the generalized Buscher rules in terms of one unambiguous parameter and some arbitrary parameters at each order of $\alpha'$. The presence of these arbitrary parameters reflects the fact that the total derivative terms in the Lagrangian are not removed. However, when working with an effective action that incorporates the Bianchi identities, field redefinitions, and the removal of total derivative terms to obtain the action in the minimal scheme \cite{Metsaev:1987zx}, then the $O(1,1)$ symmetry can determine the couplings in the minimal scheme in terms of only one unambiguous parameter at each order of $\alpha'$  \cite{Garousi:2019mca, Garousi:2023kxw, Garousi:2020gio}.

On the other hand, in the case where field redefinitions are not employed in the Lagrangian at all, the $O(1,1)$ symmetry can determine the independent couplings in the Lagrangian and all parameters in the generalized Buscher rules in terms of one unambiguous parameter and some additional arbitrary parameters compared to the previous case. These arbitrary parameters reflect the fact that no field redefinition and no integration by parts have been used to reduce the independent couplings in the Lagrangian. In other words, if one uses field redefinitions and integration by parts on the resulting action, then all arbitrary parameters can be shifted to zero or any other values. In this paper, we specifically focus on this latter scenario.

In this scenario, it becomes straightforward to find the T-duality invariant Lagrangian corresponding to the effective action in a particular scheme that is invariant under T-duality \cite{Meissner:1996sa, Garousi:2019wgz}, up to some anomalous total derivative terms in the base space. The determination of this T-duality invariant Lagrangian is based on the same generalized Buscher rules as the T-duality invariant action in that specific scheme. By understanding the generalized Buscher rules of the effective action in a specific scheme, we can determine the majority of the arbitrary parameters in the T-duality invariant Lagrangian. Consequently, we are able to identify the T-duality invariant Lagrangian that corresponds to the effective action in that scheme.

To utilize T-duality, one needs to dimensionally reduce the couplings in the Lagrangians on a circle. To perform circular reduction, we adopt the circular reduction scheme specifically designed for the massless NS-NS fields, as introduced by Maharana and Schwarz \cite{Maharana:1992my}, \ie
 \beqa
G_{\mu\nu}=\left(\matrix{\bg_{ab}+e^{\varphi}g_{a }g_{b }& e^{\varphi}g_{a }&\cr e^{\varphi}g_{b }&e^{\varphi}&}\right),\, B_{\mu\nu}= \left(\matrix{\bb_{ab}+\frac{1}{2}b_{a }g_{b }- \frac{1}{2}b_{b }g_{a }&b_{a }\cr - b_{b }&0&}\right),\,  \Phi=\bar{\phi}+\varphi/4\,.\labell{reduc}\eeqa 
The Buscher rules in this reduction represent the following transformations:
\beqa
\varphi'= -\varphi
\,\,\,,\,\,g'_{a }= b_{a }\,\,\,,\,\, b'_{a }= g_{a } \,\,\,,\,\,\bg_{ab}'=\bg_{ab} \,\,\,,\,\,\bb_{ab}'=\bb_{ab} \,\,\,,\,\,  \bar{\phi}'= \bar{\phi}\,.\labell{T2}
\eeqa
Here, $\bg_{ab}$ represents the base space metric, $\bphi$ is the base space dilaton, $\bb_{ab}$ is the base space $B$-field, and $g_a$ and $b_b$ are two vectors, while $\vp$ is a scalar in this space. The transformations mentioned above obviously form  the  $\mathbb{Z}_2$-group  $O(1,1,\mathbb{Z})$, meaning that $(\psi')'=\psi$, where $\psi$ represents any field in the base space. 
The generalized Buscher rules can have deformations involving derivatives of the base space fields such that they satisfy the same constraint $(\psi')'=\psi$.

The covariant effective Lagrangian at the leading order of $\alpha'$ in the bulk contains terms with up to second  derivatives, as follows:
\beqa
\bL^{(0)}&=& -\frac{2}{\kappa^2} e^{-2\Phi}\sqrt{-G}\,  \left(a_1 R + a_2\nabla_{\mu}\Phi \nabla^{\mu}\Phi+a_3 H^2+a_4\nabla_\mu\nabla^\mu\Phi\right)\,.\labell{S0b0}
\eeqa
where $\kappa$  is related to the $D$-dimensional Newton’s constant. The invariance of these background independent couplings under the Buscher rules \reef{T2} fixes the Lagrangian, up to an overall factor, to be \cite{Becker:2010ij}:
\beqa
\bL^{(0)}&=& -\frac{2a_1}{\kappa^2}  e^{-2\Phi}\sqrt{-G}\,  \left( R -4\nabla_{\mu}\Phi \nabla^{\mu}\Phi-\frac{1}{12} H^2+4\nabla_\mu\nabla^\mu\Phi\right)\,.\labell{S0b}
\eeqa
For $a_1=1$, the first term corresponds to the Einstein term. The aforementioned T-duality invariant Lagrangian is essentially equivalent to the standard leading-order Lagrangian, except for a total derivative term. As we consider higher orders of $\alpha'$, we anticipate that the Lagrangian will continue to exhibit invariance under $O(1,1,\mathbb{Z})$ transformations. This invariance will determine both the Lagrangian itself and the $\alpha'$-deformations of the Buscher rules. Additionally, we expect the existence of T-duality invariant Lagrangians that correspond to any known covariant effective action in the literature.

The rest of the paper is structured as follows: 
In the next section (Section 2), we employ the Bianchi identities to determine 31 even-parity independent covariant couplings with arbitrary coefficients in the Lagrangian at order $\alpha'$. In the heterotic theory at this order, there is only one odd-parity coupling with a fixed coefficient.
In Section 3, we perform a reduction on a circle and impose constraints to ensure invariance under the generalized Buscher rules. We explore all possible deformations of the Buscher rules at order $\alpha'$ for the base space fields that satisfy the $\mathbb{Z}_2$-constraint $(\psi')'=\psi$. 
In Subsection 3.1, our focus lies on the even-parity couplings, and we conduct the necessary calculations. Our findings reveal that the $O(1,1,\mathbb{Z})$ symmetry establishes 16 relations among the 31 parameters of the Lagrangian. Consequently, the T-duality invariant Lagrangian and the corresponding generalized Buscher rules involve only one unambiguous parameter and 14 arbitrary parameters.
In Subsection 3.1.1, by selecting specific relations between these arbitrary  parameters and the unambiguous parameter to align the generalized Buscher rules with those of the Metsaev-Tseytlin and Meissner actions discovered in \cite{Garousi:2019wgz,Meissner:1996sa}, we find that the corresponding Lagrangians possess one unambiguous parameter and three arbitrary parameters. Furthermore, by assigning specific values to these arbitrary parameters, we demonstrate that the resulting Lagrangians become identical to the Metsaev-Tseytlin and Meissner Lagrangians, albeit with some total derivative terms.
Subsection 3.2 is dedicated to examining the $O(1,1,\mathbb{Z})$ symmetry of the odd-parity Lagrangian in the heterotic theory.
In Section 4, we provide a brief discussion of our results. We have used the package "xAct"  \cite{Nutma:2013zea} for performing the calculations in this paper.

\section{Independent couplings}

Using the package "xAct," one finds that there are 41 covariant and gauge-invariant couplings of the massless NS-NS fields at order $\alpha'$. However, some of these couplings are interconnected through the Bianchi identities, resulting in redundant terms. By eliminating these redundancies, we identify 31 independent geometrical couplings. There are different sets for choosing these 31 couplings. All these sets are related to each other by using the Bianchi identities. The couplings in a particular set are:
\beqa
\bL_e^{(1)}&\!\!\!\!\!=\!\!\!\!\!& -\frac{2\alpha'}{\kappa^2} e^{-2\Phi}\sqrt{-G}\,\Big[ a_{1}  H_{\alpha  }{}^{\delta  \epsilon  } H^{\alpha  
\beta  \gamma  } H_{\beta  \delta  }{}^{\varepsilon  } 
H_{\gamma  \epsilon  \varepsilon  } + 
 a_{2}  H_{\alpha  \beta  }{}^{\delta  } H^{\alpha  
\beta  \gamma  } H_{\gamma  }{}^{\epsilon  \varepsilon  } 
H_{\delta  \epsilon  \varepsilon  } + 
 a_{3}  H_{\alpha  \beta  \gamma  } H^{\alpha  \beta  
\gamma  } H_{\delta  \epsilon  \varepsilon  } H^{\delta  
\epsilon  \varepsilon  }\nn\\&& + 
 a_{4}  H_{\alpha  }{}^{\gamma  \delta  } H_{\beta  
\gamma  \delta  } R^{\alpha  \beta  } + 
 a_{5}  R_{\alpha  \beta  } R^{\alpha  
\beta  } + 
 a_{6}  H_{\alpha  \beta  \gamma  } H^{\alpha  \beta  
\gamma  } R + a_{7}  R^2 + 
 a_{8}  R_{\alpha    \beta \gamma \delta  } 
R^{\alpha  \beta  \gamma  \delta  } \nn\\&&+ 
 a_{9}  H_{\beta  \gamma  \delta  } H^{\beta  \gamma  
\delta  } \nabla_{\alpha  }\nabla^{\alpha  }\Phi  + 
 a_{10}  R \nabla_{\alpha  }\nabla^{\alpha  
}\Phi  + 
 a_{11}  H_{\beta  \gamma  \delta  } H^{\beta  \gamma  
\delta  } \nabla_{\alpha  }\Phi  \nabla^{\alpha  }\Phi  + 
 a_{12}  R \nabla_{\alpha  }\Phi  
\nabla^{\alpha  }\Phi \nn\\&& + 
 a_{13}  \nabla_{\alpha  }\nabla_{\beta  }\nabla^{\beta 
 }\Phi  \nabla^{\alpha  }\Phi  + 
 a_{14}  \nabla^{\alpha  }\Phi  \nabla_{\beta  
}R_{\alpha  }{}^{\beta  } + 
 a_{15}  \nabla_{\beta  }\nabla_{\alpha  
}R^{\alpha  \beta  } + 
 a_{16}  R^{\alpha  \beta  } \nabla_{\beta  
}\nabla_{\alpha  }\Phi \nn\\&& + 
 a_{17}  \nabla_{\alpha  }\nabla^{\alpha  }\Phi  
\nabla_{\beta  }\nabla^{\beta  }\Phi  + 
 a_{18}  \nabla_{\alpha  }\Phi  \nabla^{\alpha  }\Phi  
\nabla_{\beta  }\nabla^{\beta  }\Phi  + 
 a_{19}  \nabla^{\alpha  }\Phi  \nabla_{\beta  
}\nabla^{\beta  }\nabla_{\alpha  }\Phi  + 
 a_{20}  \nabla_{\beta  }\nabla^{\beta  }\nabla_{\alpha 
 }\nabla^{\alpha  }\Phi \nn\\&& + 
 a_{21}  H_{\alpha  }{}^{\gamma  \delta  } H_{\beta  
\gamma  \delta  } \nabla^{\alpha  }\Phi  \nabla^{\beta  }\Phi 
 + 
  a_{22}  \nabla_{\alpha  }\Phi  \nabla^{\alpha  }\Phi  
\nabla_{\beta  }\Phi  \nabla^{\beta  }\Phi  + 
 a_{23}  \nabla^{\alpha  }\Phi  \nabla_{\beta  
}\nabla_{\alpha  }\Phi  \nabla^{\beta  }\Phi \nn\\&& + 
 a_{24}  H_{\alpha  }{}^{\gamma  \delta  } H_{\beta  
\gamma  \delta  } \nabla^{\beta  }\nabla^{\alpha  }\Phi  +
  a_{25}  \nabla_{\beta  }\nabla_{\alpha  }\Phi  
\nabla^{\beta  }\nabla^{\alpha  }\Phi  + 
 a_{26}  H^{\alpha  \beta  \gamma  } \nabla_{\gamma  
}\nabla_{\delta  }H_{\alpha  \beta  }{}^{\delta  } \nn\\&&+ 
 a_{27}  H^{\beta  \gamma  \delta  } \nabla^{\alpha  
}\Phi  \nabla_{\delta  }H_{\alpha  \beta  \gamma  } + 
 a_{28}  \nabla_{\alpha  }H^{\alpha  \beta  \gamma  } 
\nabla_{\delta  }H_{\beta  \gamma  }{}^{\delta  } + 
 a_{29}  H_{\alpha  }{}^{\beta  \gamma  } \nabla^{\alpha 
 }\Phi  \nabla_{\delta  }H_{\beta  \gamma  }{}^{\delta  } \nn\\&&+
  a_{30}  H^{\alpha  \beta  \gamma  } \nabla_{\delta  
}\nabla^{\delta  }H_{\alpha  \beta  \gamma  } + 
 a_{31}  \nabla_{\delta  }H_{\alpha  \beta  \gamma  } 
\nabla^{\delta  }H^{\alpha  \beta  \gamma  }\Big]\,.\labell{all}
\eeqa
The parameters $a_1, \ldots, a_{31}$ are independent of the background and cannot be determined by gauge symmetry. Among these parameters, $a_1$ and $a_8$ are unambiguous, while the remaining parameters are ambiguous as they are related by field redefinitions. Some of the ambiguous terms are also interconnected through total derivative terms. In fact, if we were to remove those terms that are related by integration by parts, we would be left with only 20 independent terms \cite{Metsaev:1987zx}. However, as mentioned in the previous section, in such a case, T-duality is only free from anomalies if we consider the combination of bulk and boundary actions to be invariant under T-duality transformations \cite{Garousi:2021cfc}. However, in this paper, our specific focus is on constraining the bulk Lagrangian alone to be invariant under T-duality transformations. Therefore, we are not permitted to eliminate the total derivative terms in equation \reef{all}. Additionally, if we were to utilize the freedom of field redefinitions, we would be left with a Lagrangian in the minimal scheme, which has 17 independent terms. However, since our objective is to discover the T-duality invariant Lagrangian for an arbitrary scheme, we do not employ the freedom of field redefinitions.

The aforementioned couplings apply to both bosonic and heterotic effective actions. However, the heterotic string theory exhibits a gauge symmetry  anomaly that can be canceled by assuming the gauge group to be $SO(32)$ or $E_8 \times E_8$, and by introducing non-standard gauge transformations and local Lorentz transformations for the $B$-field \cite{Green:1984sg}. In this paper, we focus on the case of zero gauge field. Under this assumption, the non-standard local Lorentz transformation for the $B$-field requires the field strength of the $B$-field to be the following \cite{Green:1984sg}:
\beqa
\hat{H}_{\mu\nu\alpha}(\omega)&=& H_{\mu\nu\alpha}+\frac{3}{2}\alpha'{\cO}_{\mu\nu\alpha}(\omega)\,,\labell{replace}
\eeqa
where the  Chern-Simons three-form $\cO$   is 
\beqa
\cO_{\mu\nu\alpha}(\omega)&=&\omega_{[\mu i}{}^j\prt_\nu\omega_{\alpha] j}{}^i+\frac{2}{3}\omega_{[\mu i}{}^j\omega_{\nu j}{}^k\omega_{\alpha]k}{}^i\,\,;\,\,\,\omega_{\mu i}{}^j=\prt_\mu e_\nu{}^j e^\nu{}_i-\Gamma_{\mu\nu}{}^\rho e_\alpha{}^j e^\nu{}_i\,,
\eeqa
 where $e_\mu{}^i e_\nu{}^j\eta_{ij}=G_{\mu\nu}$.  By substituting Equation \reef{replace} into Equation \reef{S0b}, the gauge symmetry of the $B$-field yields the following bulk coupling at order $\alpha'$:
\beqa
\bL^{(1)}_o&=&-\frac{2\alpha'}{\kappa^2} \sqrt{-G} \,e^{-2\Phi}\left(-\frac{1}{4}H_{\mu\nu\alpha}\cO^{\mu\nu\alpha}(\omega)\right)\,.\labell{CS}
\eeqa
This coupling is odd under parity. It is worth noting that the Bianchi identities imply that there are no other odd-parity couplings at this order. In the following, we will thoroughly investigate the invariance of the couplings given by Equations \reef{all} and \reef{CS} under $O(1,1,\mathbb{Z})$ transformations after employing circular reduction.

\section{T-duality constraint}

We now impose the constraint that the bulk effective Lagrangians  are fully  invariant under the $O(1,1,\mathbb{Z})$ transformations. In order to achieve this, we perform a reduction of the $D$-dimensional  theory on a circle with a $U(1)$ isometry, resulting in a $(D-1)$-dimensional effective Lagrangian $L_{\rm eff}(\psi)$. We then subject $L_{\rm eff}(\psi)$ to $O(1,1,\mathbb{Z})$ transformations, yielding $L_{\rm eff}(\psi')$. The $O(1,1,\mathbb{Z})$ invariance of the Lagrangian can be expressed as follows:
\beqa
L_{\rm eff}(\psi)-L_{\rm eff}(\psi')&=&0\,.\labell{TS}
\eeqa
If the effective Lagrangian and its circular reduction have the following expansions in terms of $\alpha'$:
\begin{equation}
\bL_{\rm eff}=\sum^\infty_{n=0}\frac{\alpha'^n}{n!}\bL^{(n)},\quad L_{\rm eff}=\sum^\infty_{n=0}\frac{\alpha'^n}{n!}L^{(n)},
\end{equation}
then the constraint in equation \reef{TS} can be expressed as
\begin{equation}
\sum^\infty_{n=0}\frac{\alpha'^n}{n!}L^{(n)}(\psi)-\sum^\infty_{n=0}\frac{\alpha'^n}{n!}L^{(n)}(\psi')=0\,.\labell{ss}
\end{equation}
Assuming the following generalized Buscher rules:
\begin{equation}
\psi'=\psi_0'+\sum_{n=1}^{\infty}\frac{\alpha'^n}{n!}\psi_n',
\end{equation}
where $\psi_0'$ represents the Buscher rules as given in \reef{T2}, and $\psi_n'$ denotes the corrections at order $\alpha'^n$. Expanding the second term in \reef{ss} around the Buscher transformations $\psi_0'$, that is,
\begin{equation}
L^{(n)}(\psi')=L^{(n)}(\psi_0')+\sum_{m=1}^{\infty}\frac{\alpha'^m}{m!}L^{(n,m)}(\psi_0')\,,
\end{equation}
we obtain the following constraint:
\begin{equation}
\sum^\infty_{n=0}\frac{\alpha'^n}{n!}L^{(n)}(\psi)-\sum^\infty_{n=0}\frac{\alpha'^n}{n!}L^{(n)}(\psi_0')-\sum^\infty_{n=0,m=1}\frac{\alpha'^{n+m}}{n!m!}L^{(n,m)}(\psi_0')=0\,. \labell{TSn}
\end{equation}
To determine the appropriate constraints on the effective Lagrangians, each term at every order of $\alpha'$ must be set to zero.

The constraint given by  \reef{TSn} at the leading order is:
\begin{equation}
L^{(0)}(\psi)-L^{(0)}(\psi'_0) = 0\,, \labell{TS0}
\end{equation}
where $L^{(0)}(\psi)$ represents the circular reduction of the leading-order Lagrangian, and $L^{(0)}(\psi'_0)$ is its transformation under the Buscher rules given by  \reef{T2}. The reduction of the leading-order Lagrangian, as expressed in \reef{S0b}, is \cite{Kaloper:1997ux}
\beqa
L^{(0)}(\psi)&=& -\frac{2}{\kappa'^2} e^{-2\bphi}\sqrt{-\bg}\,  \Big[ \bar{R} -4\nabla_{a}\bphi \nabla^{a}\bphi-\frac{1}{12} \bH^2+4\nabla_a\nabla^a\bphi\nn\\&&\qquad\qquad\qquad\qquad\qquad
-\frac{1}{4}(e^{\vp}V^2+e^{-\vp}W^2)-\frac{1}{4}\nabla_a\vp\nabla^a\vp\Big]\,.\labell{S0r}
\eeqa
where $\kappa'$  is related to the ($D$-1)-dimensional Newton’s constant. In the given equation, the curvature and covariant derivatives on the right-hand side are constructed using the metric of the base space. As for the variables $V$, $W$, and $\bH$, they are defined as follows:
\beqa
V_{ab}&=&\prt_a g_b-\prt_b g_a\,,\nn\\
W_{ab}&=&\prt_a b_b-\prt_b b_a\,,\nn\\
\bH_{abc}&=&3\prt_{[a}\bb_{bc]}-\frac{3}{2}g_{[a}W_{bc]}-\frac{3}{2}b_{[a}V_{bc]}\,.
\eeqa
 Since $\bH$ is not the exterior derivative of a two-form, it satisfies the following anomalous Bianchi identity \cite{Kaloper:1997ux}:
 \beqa
 \prt_{[a} \bH_{bcd]}&=&-\frac{3}{2}V_{[ab}W_{cd]}\,.\labell{anB}
 \eeqa
  Our notation for antisymmetry is such that, for example, $g_{[a}W_{bc]}=\frac{1}{3}(g_aW_{bc}-g_{b}W_{ac}-g_cW_{ba})$. It is evident that the reduction given by Equation \reef{S0r} satisfies the T-duality constraint in Equation \reef{TS0}.

The constraint \reef{TSn} at order $\alpha'$  is
\beqa
L^{(1)}(\psi)-L^{(1)}(\psi'_0)-L^{(0,1)}(\psi'_0)&=& 0\,,\labell{TS11}
\eeqa
where $L^{(0,1)}(\psi'_0)$ is the Taylor expansion of the reduction \reef{S0r} at first order.  Writing the first order corrections to the Buscher rules as 
\beqa
&&\varphi'= -\varphi+\alpha'\Delta\vp,\,\,g'_{a }= b_{a }+\alpha'e^{\vp/2}\Delta g _a,\,\, b'_{a }= g_{a }+\alpha'e^{-\vp/2}\Delta b _a,\nn\\&&\bH_{abc}'=\bH_{abc}+\alpha'\Delta\bH_{abc},\,\, 
\bphi'=\bphi+\alpha'\Delta\bphi,\,\, \bg_{ab}'=\bg_{ab}+\alpha'\Delta \bg_{ab}\,,\labell{T22}
\eeqa
  one finds  $L^{(0,1)}(\psi'_0)$ to be 
\beqa
L^{(0,1)}&\!\!\!\!\!=\!\!\!\!\!& -\frac{2\alpha' }{\kappa'^2} e^{-2\bphi}\sqrt{-\bg} \,  \Big\{ 
\frac{1}{4}\Big(  e^\vp V^2-e^{-\vp}W^2\Big)\Delta\vp+\frac{1}{2}\nabla_a\vp\nabla^a(\Delta\vp)-\frac{1}{6}\bH^{abc}\Delta\bH_{abc}\nn\\
 &&
+e^{-\vp}W_{ab}\nabla^b(e^{\vp/2}\Delta g^a)+e^{\vp}V_{ab}\nabla^b(e^{\vp/2}\Delta b^a)-8\nabla_a\bphi\nabla^a(\Delta\bphi)+4\nabla_a\nabla^a(\Delta\bphi)\nn\\&&
(\frac{1}{2}\Delta\bg^a{}_a\!-\!2\Delta \bphi )\Big[ \bar{R}\! -4\nabla_{a}\bphi \nabla^{a}\bphi-\frac{1}{12} \bH^2+4\nabla_a\nabla^a\bphi
-\frac{1}{4}(e^{\vp}V^2+e^{-\vp}W^2)-\frac{1}{4}\nabla_a\vp\nabla^a\vp\Big] \nn\\&&
-\Delta\bg^{ab}\Big[\bar{R}_{ab}-\frac{1}{4}\bH^2_{ab}-\frac{1}{2}(e^{\vp}V^2_{ab}+e^{-\vp}W^2_{ab})+4\nabla_b\nabla_a\bphi-4\nabla_a\bphi\nabla_b\bphi-\frac{1}{4}\nabla_a\vp\nabla_b\vp\Big]\nn\\&&+2\nabla^a\bphi\nabla_a(\Delta\bg^b{}_b)-4\nabla^a\bphi\nabla_b(\Delta\bg_a{}^b)+\nabla_b\nabla_a(\Delta\bg^{ab})-\nabla_b\nabla^b(\Delta\bg^a{}_a)\Big\}\,.\labell{delS}
 \eeqa
 As the generalized Buscher rules are required to form the $\mathbb{Z}_2$-group, the corrections must satisfy the following relations \cite{Garousi:2019wgz}:
\beqa
\Delta\bphi (\psi)+\Delta\bphi (\psi_0') &=&0\,,\nn\\
\Delta \bg_{ab} (\psi)+\Delta \bg_{ab} (\psi'_0) &=&0\,,\nn\\
-\Delta\vp (\psi)+\Delta\vp (\psi_0') &=&0\,,\nn\\
\Delta b_a (\psi)+\Delta g_a (\psi'_0) &=&0\,,\nn\\
\Delta g_a (\psi)+\Delta b_a (\psi'_0)&=&0\,,\nn\\
\Delta \bH_{abc} (\psi)+\Delta \bH_{abc} (\psi'_0) &=&0\,.\labell{Z22}
\eeqa
The correction $\Delta \bH_{abc}$ is related to the corrections $\Delta g_a$ and $\Delta b_a$ through the following relation, which arises from the Bianchi identity in Equation \reef{anB}:
\beqa
\Delta\bH_{abc}&=&\tilde H_{abc}-3e^{-\vp/2}W_{[ab}\Delta b_{c]}-3e^{\vp/2}\Delta g_{[a}V_{bc]}\,.
\eeqa
Here, $\tilde{H}_{abc}$ represents a $U(1) \times U(1)$ gauge invariant exact 3-form, meaning $\tilde{H} = d\tilde{B}$, where $\tilde{B}$ is a $U(1) \times U(1)$ gauge invariant 2-form at order $\alpha'$ \cite{Kaloper:1997ux,Garousi:2023kxw}. This 2-form, along with the corrections $\Delta\bphi,\Delta\bg_{ab},\Delta \varphi$, $\Delta g_a$, and $\Delta b_a$, contains contractions involving the base space fields  at order $\alpha'$.

While in the bosonic theory, $L^{(1)}$ consists only of even-parity terms, in the heterotic theory, $L^{(1)}$ includes both even and odd parity terms. Consequently, $L^{(1)}$ and the corrections to the Buscher rules appearing in $L^{(0,1)}(\psi'_0)$ contain both even and odd parity terms. In the next subsection, we will analyze the even-parity terms.

\subsection{Even-parity couplings}

By utilizing the reductions given in Equation \reef{reduc}, it is straightforward to determine the circular reduction of Equation \reef{all} and obtain $L^{(1)}(\psi)$. We can then transform $L^{(1)}(\psi)$ under the Buscher rules in Equation \reef{T2} to obtain $L^{(1)}(\psi_0')$. 

To calculate the corresponding even-parity contribution from $L^{(0,1)}(\psi'_0)$, we need to write $\Delta\bphi$, $\Delta\bg_{ab}$, $\Delta \varphi$, $\Delta g_a$, $\Delta b_a$, and $\tilde{B}$ in \reef{delS} as all possible contractions of the base space fields $\bR$, $\bH$, $W$, $V$, $\nabla\vp$, and $\nabla\bphi$ at order $\alpha'$, which satisfy the constraint \reef{Z22}. The corrections $\Delta\bphi$, $\Delta\bg_{ab}$, $\Delta \varphi$, and $\Delta b_a$ must have even-parity terms, while the corrections $\Delta g_a$ and $\tilde{B}$ must have odd-parity terms. In other words, one should take into account the following corrections to the Buscher rules:
 \beqa
 \Delta\bphi&=&f_3(e^{\vp}V^2-e^{-\vp}W^2)+f_9\nabla_a\nabla^a\vp+f_6\nabla_a\vp\nabla^a\vp\,,\nn\\
 \Delta\vp&=&e_1\bH^2+e_2\bar{R}+e_3(e^{\vp}V^2+e^{-\vp}W^2)+e_8\nabla_a\nabla^a\bphi+e_5\nabla_a\bphi\nabla^a\bphi+e_7\nabla_a\vp\nabla^a\vp\,,\nn\\
 \Delta\bg_{ab}&=&d_5(e^{\vp}V^2_{ab}-e^{-\vp}W^2_{ab})+d_6\bg_{ab}(e^{\vp}V^2-e^{-\vp}W^2)+d_{11}(\nabla_a\vp\nabla_b\bphi+\nabla_b\vp\nabla_a\bphi)\nn\\&&+d_{12}\bg_{ab}\nabla_a\vp\nabla^c\bphi+d_{17}\nabla_b\nabla_a\vp+d_{18}\bg_{ab}\nabla_c\nabla^c\vp\,,\nn\\
\Delta g_a &=&b_1e^{\vp/2}\bH_{abc} V^{bc}+b_3\nabla_b(e^{-\vp/2} W_a{}^{b})+b_2e^{-\vp/2} W_{ab}\nabla^b\bphi+b_4e^{-\vp/2}W_{ab}\nabla^b\vp\,,\nn\\
\Delta b_a &=&-b_1e^{-\vp/2}\bH_{abc} W^{bc}-b_3\nabla_b(e^{\vp/2} V_a{}^{b})-b_2e^{\vp/2}V_{ab}\nabla^b\bphi+b_4e^{\vp/2} V_{ab}\nabla^b\vp\,,\nn\\
\tilde B_{ab}&=&aa_1(W_{a}{}^cV_{bc}-W_{b}{}^cV_{ac})+aa_4\bH_{abc}\nabla^c\vp\,.\labell{tH}
\eeqa
Here, $f_3, \cdots, aa_4$ are 21  parameters that can be determined by the constraint given in Equation \reef{TS11}. It should be noted that the mentioned deformations satisfy the $\mathbb{Z}_2$-constraint in Equation \reef{Z22}. When these deformations are inserted into Equation \reef{delS}, it can be observed that $L^{(0,1)}(\psi'_0)$ yields even-parity terms.

The constraint given by Equation \reef{TS11} then produces 37 relations between the parameters in the deformations and the parameters in the Lagrangian \reef{all} as follows:
\beqa
&&a_{20} = 2 a_{15}, 
 a_{21} = -48 a_{1} + 16 a_{2} + a_{10}/2 + a_{12}/2 - a_{17}/4 - a_{18}/8, \nn\\&&
 a_{22} = 2 a_{10} + 2 a_{12} - 2 a_{17} - (3 a_{18})/2, 
 a_{23} = -8 a_{10} - 4 a_{13} + 4 a_{14} + 8 a_{17} + 2 a_{18}, \nn\\&&
 a_{25} = 2 a_{10} + 2 a_{12} + a_{16} - a_{17} - a_{18}/2 + a_{19}, 
 a_{27} = -6 a_{11} - (9 a_{12})/2 + (3 a_{13})/2 - (9 a_{14})/4 \nn\\&&+ 
   a_{16}/4 - a_{17} + a_{18}/2 - a_{19}/2 + a_{24}, 
 a_{28} = -12 a_{1} + 4 a_{2}, a_{29} = 48 a_{1} - 16 a_{2}, \nn\\&&
 a_{30} = 8 a_{1} + (4 a_{2})/3 - 
   12 a_{3} + (11 a_{10})/96 + (5 a_{11})/4 + (91 a_{12})/96 - (7 a_{13})/
    24 + (5 a_{14})/12\nn\\&& - 
   a_{16}/48 + (5 a_{17})/32 - (41 a_{18})/384 + (5 a_{19})/48, 
 a_{31} = 4 a_{1} + (4 a_{2})/3 - 
   12 a_{3} + (11 a_{10})/96 \nn\\&&+ \!(5 a_{11})/4 \!+ \!(91 a_{12})/96 \!- \!(7 a_{13})/
    24 \!+\! (5 a_{14})/12 \!- \!
   a_{16}/48\! + \!(5 a_{17})/32 \!- \!(41 a_{18})/384 \!+ \!(5 a_{19})/48, \nn\\&&
 a_{4} = -24 a_{1} - 4 a_{2} - a_{16}/16, a_{5} = a_{16}/4, 
 a_{6} = -((7 a_{10})/48) - a_{11}/4 - (5 a_{12})/24 + a_{13}/24\nn\\&& - 
   a_{14}/12 - a_{15}/12 + a_{17}/48 + (5 a_{18})/192 - a_{19}/48, 
 a_{7} = (3 a_{10})/8 + a_{12}/8 - a_{17}/8 - a_{18}/32, \nn\\&&a_{8} = 24 a_{1}, 
 a_{9} = -(a_{10}/12) - a_{11} - (5 a_{12})/12 + a_{13}/12 - a_{14}/6 - 
   a_{15}/3 - a_{17}/6 - a_{19}/12, \nn\\&&aa_{1} = -144 a_{1}, 
 aa_{4} = -144 a_{1} - 
   24 a_{2} - (3 a_{10})/2 - (3 a_{13})/8 + (3 a_{14})/4 - (3 a_{16})/
    8 + (3 a_{17})/4\nn\\&& - 3 a_{24} - 6 a_{26}, b_{1} = -12 a_{1} - a_{26}, 
 b_{2} = -96 a_{1} + 32 a_{2} + 2 a_{24}, 
 b_{3} = 96 a_{1} - 8 a_{2} + a_{16}/8 + 2 a_{26}, \nn\\&&
 b_{4} = 24 a_{1} + 8 a_{2} + a_{10}/2 + a_{13}/8 - a_{14}/4 - a_{17}/4 + 
   a_{24}/2, d_{11} = a_{10} + a_{12} - a_{17}/2 - a_{18}/4, \nn\\&&
 d_{12} = -2 a_{12} + a_{13}/2 - a_{14} - 2 a_{15} - a_{17} - a_{19}/2, 
 d_{17} = -a_{10} - a_{13}/4 + a_{14}/2 + a_{17}/2, \nn\\&&
 d_{18} = a_{10} + a_{12} - a_{13}/4 + a_{14}/2 - a_{17}/2 - a_{18}/4, 
 d_{5} = 24 a_{1} - 8 a_{2} + a_{16}/8, \nn\\&&
 d_{6} = -18 a_{3} + (11 a_{10})/64 + (3 a_{11})/8 + (19 a_{12})/64 - 
   a_{13}/16 + a_{14}/8 + a_{15}/8 - a_{17}/64 \nn\\&&- (9 a_{18})/256 + a_{19}/32,
  e_{1} = -12 a_{3} + a_{10}/32 + a_{11}/4 + (19 a_{12})/96 - a_{13}/16 + 
   a_{14}/12 + a_{17}/32 \nn\\&&- (3 a_{18})/128 + a_{19}/48, 
 e_{2} = a_{10} + a_{13}/4 + a_{15} - a_{17}/2, 
 e_{3} = 36 a_{1} - 4 a_{2} - 
   18 a_{3} - (5 a_{10})/64  \nn\\&&+ (3 a_{11})/8 + (19 a_{12})/64- a_{13}/8 + 
   a_{14}/8 - a_{15}/8 + a_{16}/16 + (7 a_{17})/64 - (9 a_{18})/256 + 
   a_{19}/32,  \nn\\&&
 e_{5} = 8 a_{12}\! -\! 3 a_{13} \!+\! 4 a_{14} \!+\! 2 a_{17} \!-\! a_{18} \!+\! a_{19}, 
 e_{7} = 48 a_{1}\! +\! a_{13}/16 \!-\! a_{14}/4 \!+\! a_{17}/8 \!+\! a_{18}/16 \!+\! a_{19}/16, \nn\\&&
  e_{8} = -4 a_{12} + 2 a_{13} - 2 a_{14} + 4 a_{15} + a_{18}, 
 f_{3} = 6 a_{1} - 
   2 a_{2} - (207 a_{3})/2 + (253 a_{10})/256 \nn\\&& + (57 a_{11})/
    32 + (365 a_{12})/256 - (17 a_{13})/64 + (19 a_{14})/32 + (23 a_{15})/
    32 + a_{16}/32 - (39 a_{17})/256  \nn\\&&- (175 a_{18})/1024 + (19 a_{19})/
    128, f_{6} = -((23 a_{12})/2) + (23 a_{13})/8 - (23 a_{14})/
    4 - (23 a_{15})/2 - 6 a_{17}  \nn\\&&- a_{18}/8 - (23 a_{19})/8, 
 f_{9} = 6 a_{10} + 6 a_{12} - (3 a_{13})/2 + 3 a_{14} - 
   3 a_{17} - (3 a_{18})/2\,.\labell{abef}
\eeqa
By inserting the aforementioned relations into the Lagrangian \reef{all} and the $\alpha'$-correction of the Buscher rules \reef{tH}, one can derive the Lagrangian and the corresponding generalized Buscher rules in terms of one unambiguous parameter, $a_1$, as well as 14 arbitrary parameters:
\beqa
a_{10}, a_{11}, a_{12}, a_{13}, a_{14}, a_{15}, a_{16}, a_{17}, a_{18}, a_{19}, a_2, a_{24}, a_{26}, a_3 \,.\label{a103}
\eeqa
For any choice of these arbitrary parameters in terms of the unambiguous parameter $a_1$, one can find the T-duality invariant Lagrangian and its corresponding generalized Buscher rules within a specific scheme. It is evident that the coefficient of the first term in $\tilde{B}_{ab}$ serves as the unambiguous parameter, which cannot be set to zero. Furthermore, for any selection of the arbitrary parameters, it is not possible to simultaneously set all deformations $\Delta\varphi$, $\Delta\bar{\phi}$, $\Delta\bar{g}_{ab}$, $\Delta g_a$, and $\Delta b_a$ to zero. However, it is feasible to choose certain arbitray parameters such that some of them are set to zero. For example, by setting the following values for the arbitrary parameters:
\beqa
&&a_{11}=-32a_1+\frac{a_{10}}{12}, a_{12}=-a_{10},  a_{14}=\frac{a_{13}}{2}, a_{15}=96a_1-\frac{a_{13}}{4},  a_{17}=2(96a_1+a_{10}),  \nn\\&&a_{18}=-4(96a_1+a_{10}), a_{19}=-768a_1+a_{13},  a_2=9a_1+\frac{a_{16}}{64},  a_{24}=-96a_1-\frac{a_{16}}{4}, \nn\\&& a_{26}=-12a_1,  a_3=\frac{a_{10}-864a_1}{1152}\,, \label{aaa}
\eeqa
one finds $\Delta\varphi=\Delta\bar{\phi}=\Delta g_a=\Delta b_a=0$ and
\beqa
\Delta\bar{g}_{ab}&=-48a_1(e^{\varphi}V^2_{ab}-e^{-\varphi}W^2_{ab}-2\nabla_b\nabla_a\varphi)\,, \nn \\
\tilde{B}_{ab}&=-144a_1(W_a{}^cV_{bc}-V_a{}^cW_{bc}-\bar{H}_{abc}\nabla^c\varphi)\,. 
\eeqa
By inserting the values \reef{aaa} into \reef{all}, one can find its corresponding T-duality invariant Lagrangian in terms of the unambiguous parameter $a_1$ and three arbitrary parameters $a_{10}$, $a_{13}$, and $a_{16}$.

A connection can be established between the arbitrary parameters \reef{a103} and the unambiguous parameter $a_1$ by comparing the generalized Buscher rules with existing ones in the literature. By imposing these relationships between the arbitrary parameters, the T-duality invariant Lagrangian corresponding to those generalized Buscher rules can be derived. However, it turns out that the number of these relationships is less than 14, which means that the resulting T-duality invariant Lagrangian will still have some arbitrary parameters. Since there are generalized Buscher rules in the literature associated with the Metsaev-Tseytlin action and the Meissner action, we will further investigate this in the subsequent subsection.

\subsubsection{Comparing with the Metsaev-Tseytlin action}

It has been shown in \cite{Garousi:2019wgz} that the effective action at order $\alpha'$ in the Metsaev-Tseytlin scheme is invariant under the $O(1,1,\mathbb{Z})$ transformation, except for some anomalous total derivative terms in the base space. The corresponding generalized Buscher rules are as follows:
 \beqa
  \Delta \bar{g}_{ab}&=&48a_1\Big(e^\vp V_a {}^c V_{bc}-e^{-\vp}W_a {}^c W_{bc}\Big) \,,\nn\\
  \Delta\bphi&=&12a_1\Big(e^\vp V^2- e^{-\vp}W^2\Big) \,,\nn\\
  \Delta\vp&=&48a_1\Big(\nabla_a\vp\nabla^a\vp+e^\vp V^2+e^{-\vp}W^2\Big)\,, \nn\\
  \Delta g_{a}&=&24a_1\Big(2e^{-\vp/2}\nabla^b W_{ab}+e^{\vp/2}\bH_{abc} V^{bc}-4e^{-\vp/2}\nabla^b\bphi W_{ab}\Big)\,,\nn\\
   \Delta b_{a}&=&-24a_1\Big(2e^{\vp/2}\nabla^b V_{ab}+e^{-\vp/2}\bH_{abc} W^{bc}-4e^{\vp/2}\nabla^b\bphi V_{ab}\Big)\,,\nn\\
   \tilde B_{ab}&=&-144a_1(W_{a}{}^cV_{bc}-W_{b}{}^cV_{ac})\,.\labell{dbH1}
  \eeqa
Comparing the generalized Buscher rules \reef{tH} in which the relations \reef{abef} are inserted, with the above deformations, one finds the following relation for the 14 arbitrary parameters:
\beqa
&&a_{12}=-a_{10},a_{14}=a_{13}/2,a_{15}=-a_{13}/4,a_{17}=2a_{10},a_{18}=-4a_{10},a_{19}=a_{13},\nn\\&&a_2=-3a_1+a_{16}/64,a_{24}=48a_1-a_{16}/4,a_{26}=-36a_1,a_3=(24a_{11}-a_{10})/1152\,.
\eeqa
There are still  four  arbitrary parameters $a_{10}, a_{11}, a_{13}$,  and $a_{16}$. If one chooses these parameters to be zero, then the T-duality invariant Lagrangian becomes
\beqa
\bL_e^{(1)}&\!\!\!\!\!=\!\!\!\!\!& -\frac{48a_1\alpha'}{\kappa^2} e^{-2\Phi}\sqrt{-G}\,\Big[ \frac{1}{24} H_{\alpha }{}^{\delta \epsilon } H^{\alpha \beta 
\gamma } H_{\beta \delta }{}^{\varepsilon } H_{\gamma \epsilon 
\varepsilon } -  \frac{1}{8} H_{\alpha \beta }{}^{\delta } 
H^{\alpha \beta \gamma } H_{\gamma }{}^{\epsilon \varepsilon } 
H_{\delta \epsilon \varepsilon } -  \frac{1}{2} H_{\alpha }{}^{
\gamma \delta } H_{\beta \gamma \delta } R^{\alpha 
\beta } \nn\\&&+  R_{\alpha \beta\gamma  \delta } 
R^{\alpha \beta \gamma \delta } - 4 H_{\alpha 
}{}^{\gamma \delta } H_{\beta \gamma \delta } \nabla^{\alpha }
\Phi \nabla^{\beta }\Phi + 2 H_{\alpha }{}^{\gamma \delta } 
H_{\beta \gamma \delta } \nabla^{\beta }\nabla^{\alpha }\Phi 
-  \frac{3}{2} H^{\alpha \beta \gamma } \nabla_{\gamma 
}\nabla_{\delta }H_{\alpha \beta }{}^{\delta } \nn\\&&+ 2 H^{\beta 
\gamma \delta } \nabla^{\alpha }\Phi \nabla_{\delta 
}H_{\alpha \beta \gamma } \!- \! \nabla_{\alpha }H^{\alpha \beta 
\gamma } \nabla_{\delta }H_{\beta \gamma }{}^{\delta } \!+ \!4 
H_{\alpha }{}^{\beta \gamma } \nabla^{\alpha }\Phi 
\nabla_{\delta }H_{\beta \gamma }{}^{\delta } \!+ \!\frac{1}{6} 
H^{\alpha \beta \gamma } \nabla_{\delta }\nabla^{\delta 
}H_{\alpha \beta \gamma }\Big].\labell{ST1}
\eeqa
If one chooses the unambiguous parameters to be $24 a_1=-\lambda_0$, and uses the following identity:
\beqa
H_\alpha{}^{\gamma\delta}H_{\beta\gamma\delta}R^{\alpha\beta}-H_\alpha{}^{\delta\epsilon}H^{\alpha\beta\gamma}R_{\beta\gamma\delta\epsilon}+H^{\alpha\beta\gamma}\nabla_\gamma\nabla_\delta H_{\alpha\beta}{}^\delta-\frac{1}{3}H^{\alpha\beta\gamma}\nabla_\delta\nabla^\delta H_{\alpha\beta\gamma}&=&0.
\eeqa
Then, up to some total derivative terms, the Lagrangian becomes the one in the Metsaev-Tseytlin action \cite{Metsaev:1987zx}. In fact, one can write the above Lagrangian as
\beqa
\bL_e^{(1)}&=&\bL_{MT}^{(1)} +\frac{2\lambda_0\alpha' }{\kappa^2} \nabla_\alpha(e^{-2\Phi}J^\alpha)\,,\labell{totMT}
\eeqa
where the current in the total derivative is given by
\beqa
J^{\alpha} = 2 H^{\alpha \gamma \delta } H_{\beta \gamma \delta } \nabla^{\beta }\Phi -  H^{\alpha \beta \gamma } \nabla_{\delta}H_{\beta \gamma }{}^{\delta },
\eeqa
and the Metsaev-Tseytlin Lagrangian is
\beqa
\bL_{MT}^{(1)}&=& \frac{2\lambda_0\alpha'}{\kappa^2} e^{-2\Phi}\sqrt{-G}\,\Big[ 
R_{\alpha  \beta \gamma \delta } R^{\alpha 
\beta \gamma \delta }  -\frac{1}{2} H_{\alpha }{}^{\delta \epsilon } 
H^{\alpha \beta \gamma } R_{\beta \gamma\delta  
\epsilon }  +\frac{1}{24}H_{\alpha }{}^{\delta \epsilon } H^{\alpha \beta \gamma } 
H_{\beta \delta }{}^{\varepsilon } H_{\gamma \epsilon 
\varepsilon } \nn\\&&\qquad\qquad\qquad\qquad\qquad\quad- \frac{1}{8} H_{\alpha \beta 
}{}^{\delta } H^{\alpha \beta \gamma } H_{\gamma }{}^{\epsilon 
\varepsilon } H_{\delta \epsilon \varepsilon } \Big]\,.\labell{fs3}
\eeqa
It is important to note that the Metsaev-Tseytlin Lagrangian is derived from the S-matrix elements, and the total derivative terms cannot be fixed by the S-matrix method. The above calculation indicates that, up to certain total derivative terms, the Metsaev-Tseytlin Lagrangian is indeed invariant under T-duality transformations.  The unambiguous parameter in the Lagrangian and the generalized Buscher rules is $\lambda_0=-\frac{1}{4}$ for bosonic string theory and $\lambda_0=-\frac{1}{8}$ for the heterotic theory.

\subsubsection{Comparing with the Meissner action}

It has been shown in \cite{Meissner:1996sa} that using field redefinitions and adding total derivative terms to the Metsaev-Tseytlin action, one can write this action in another scheme that its cosmological reduction has only first time derivative. The Lagrangian in the Meissner scheme is \cite{Meissner:1996sa}
\beqa
\bL_M^{(1)}&=&\frac{2\lambda_0\alpha' }{\kappa^2}\sqrt{-G} e^{-2\Phi}\Big[R_{GB}^2+\frac{1}{24} H_{\alpha }{}^{\delta \epsilon } H^{\alpha \beta
\gamma } H_{\beta \delta }{}^{\varepsilon } H_{\gamma \epsilon
\varepsilon } -  \frac{1}{8} H_{\alpha \beta }{}^{\delta }
H^{\alpha \beta \gamma } H_{\gamma }{}^{\epsilon \varepsilon }
H_{\delta \epsilon \varepsilon }\nn\\&&  + \frac{1}{144} H_{\alpha
\beta \gamma } H^{\alpha \beta \gamma } H_{\delta \epsilon
\varepsilon } H^{\delta \epsilon \varepsilon }+ H_{\alpha }{}^{
\gamma \delta } H_{\beta \gamma \delta } R^{\alpha
\beta } -  \frac{1}{6} H_{\alpha \beta \gamma } H^{\alpha \beta
\gamma } R  -
\frac{1}{2} H_{\alpha }{}^{\delta \epsilon } H^{\alpha \beta
\gamma } R_{\beta \gamma \delta \epsilon }\nn\\&& -
\frac{2}{3} H_{\beta \gamma \delta } H^{\beta \gamma \delta }
\nabla_{\alpha }\nabla^{\alpha }\Phi + \frac{2}{3} H_{\beta
\gamma \delta } H^{\beta \gamma \delta } \nabla_{\alpha }\Phi
\nabla^{\alpha }\Phi + 8 R \nabla_{\alpha }\Phi
\nabla^{\alpha }\Phi + 16 \nabla_{\alpha }\Phi \nabla^{\alpha
}\Phi \nabla_{\beta }\nabla^{\beta }\Phi \nn\\&&- 16
R_{\alpha \beta } \nabla^{\alpha }\Phi \nabla^{\beta
}\Phi - 16 \nabla_{\alpha }\Phi \nabla^{\alpha }\Phi \nabla_{
\beta }\Phi \nabla^{\beta }\Phi + 2 H_{\alpha }{}^{\gamma
\delta } H_{\beta \gamma \delta } \nabla^{\beta
}\nabla^{\alpha }\Phi \Big]\,,\labell{mie}
\eeqa
where $R_{GB}^2=R_{\alpha\beta\mu\nu}R^{\alpha\beta\mu\nu}-4R_{\alpha\beta}R^{\alpha\beta}+R^2$ is the Gauss-Bonnet couplings.
It has been observed in  \cite{Kaloper:1997ux} that the circular reduction of the above action is invariant under the $O(1,1,\MZ)$ transformations with the following $\alpha'$-correction to the Buscher rules:  
 \beqa
   \Delta \bar{g}_{ab}&=&0\,, \nn\\
  \Delta\bphi&=&0\,, \nn\\
  \Delta\vp&=&-\lambda_0\Big(2\nabla_a\vp\nabla^a\vp+e^\vp V^2+e^{-\vp}W^2\Big)\,, \nn\\
  \Delta g_{a}&=&-\lambda_0\Big(2e^{-\vp/2}\nabla^b\vp W_{ab}+e^{\vp/2}\bH_{abc} V^{bc} \Big)\,,\nn\\
   \Delta b_{a}&=&-\lambda_0\Big(2e^{\vp/2}\nabla^b\vp V_{ab}-e^{-\vp/2}\bH_{abc} W^{bc} \Big)\,,\nn\\
  \tilde B_{ab}&=&12\lambda_0W_{[a}{}^c V_{b]c}\,.\labell{dbH2}
  \eeqa
Comparing the generalized Buscher rules \reef{tH} in which the relations \reef{abef} are inserted, with the above deformations for $24a_1=-\lambda_0$, one finds the following relations for the 14 arbitrary parameters:
\beqa
&&a_{11}=a_{10}/12,a_{12}=-a_{10},a_{14}=a_{13}/2,a_{15}=-a_{13}/4,a_{17}=2a_{10},a_{18}=-4a_{10},a_{19}=a_{13},\nn\\&&a_2=3a_1+a_{16}/64,a_{24}=-a_{16}/4,a_{26}=-36a_1,a_3=a_{10}/1152\,.
\eeqa
There are still  three arbitrary parameters $a_{10}, a_{13},$ and $a_{16}$. If one chooses these parameters to be the following
\beqa
a_{10}=192a_1,a_{13}=0,a_{16}=-384a_1\,,
\eeqa
then the T-duality invariant Lagrangian becomes
\beqa
\bL_e^{(1)}&\!\!\!\!\!=\!\!\!\!\!& \frac{2\lambda_0\alpha'}{\kappa^2} e^{-2\Phi}\sqrt{-G}\,\Big[\frac{1}{24} H_{\alpha }{}^{\delta \epsilon } H^{\alpha \beta 
\gamma } H_{\beta \delta }{}^{\varepsilon } H_{\gamma \epsilon 
\varepsilon } -  \frac{1}{8} H_{\alpha \beta }{}^{\delta } 
H^{\alpha \beta \gamma } H_{\gamma }{}^{\epsilon \varepsilon } 
H_{\delta \epsilon \varepsilon } + \frac{1}{144} H_{\alpha 
\beta \gamma } H^{\alpha \beta \gamma } H_{\delta \epsilon 
\varepsilon } H^{\delta \epsilon \varepsilon }\nn\\&& + \frac{1}{2} 
H_{\alpha }{}^{\gamma \delta } H_{\beta \gamma \delta } 
R^{\alpha \beta } - 4 R_{\alpha \beta } 
R^{\alpha \beta } -  \frac{1}{6} H_{\alpha \beta 
\gamma } H^{\alpha \beta \gamma } R+ R^2 + 
R_{\alpha  \beta \gamma\delta } R^{\alpha 
\beta \gamma \delta } -  \frac{2}{3} H_{\beta \gamma \delta } 
H^{\beta \gamma \delta } \nabla_{\alpha }\nabla^{\alpha }\Phi 
\nn\\&&+ 8 R\nabla_{\alpha }\nabla^{\alpha }\Phi + 
\frac{2}{3} H_{\beta \gamma \delta } H^{\beta \gamma \delta } 
\nabla_{\alpha }\Phi \nabla^{\alpha }\Phi \!- \!8 R
\nabla_{\alpha }\Phi \nabla^{\alpha }\Phi \!-\! 16 
R^{\alpha \beta } \nabla_{\beta }\nabla_{\alpha }\Phi 
\!+\! 16 \nabla_{\alpha }\nabla^{\alpha }\Phi \nabla_{\beta 
}\nabla^{\beta }\Phi \nn\\&&- 32 \nabla_{\alpha }\Phi \nabla^{\alpha 
}\Phi \nabla_{\beta }\nabla^{\beta }\Phi\! - \!4 H_{\alpha 
}{}^{\gamma \delta } H_{\beta \gamma \delta } \nabla^{\alpha }
\Phi \nabla^{\beta }\Phi \!+ \!16 \nabla_{\alpha }\Phi 
\nabla^{\alpha }\Phi \nabla_{\beta }\Phi \nabla^{\beta }\Phi 
\!+\! 4 H_{\alpha }{}^{\gamma \delta } H_{\beta \gamma \delta } 
\nabla^{\beta }\nabla^{\alpha }\Phi\nn\\&& - 16 \nabla_{\beta 
}\nabla_{\alpha }\Phi \nabla^{\beta }\nabla^{\alpha }\Phi -  
\frac{3}{2} H^{\alpha \beta \gamma } \nabla_{\gamma }\nabla_{
\delta }H_{\alpha \beta }{}^{\delta } -  \nabla_{\alpha 
}H^{\alpha \beta \gamma } \nabla_{\delta }H_{\beta \gamma 
}{}^{\delta } + 4 H_{\alpha }{}^{\beta \gamma } \nabla^{\alpha 
}\Phi \nabla_{\delta }H_{\beta \gamma }{}^{\delta } \nn\\&&+ 
\frac{1}{2} H^{\alpha \beta \gamma } \nabla_{\delta }\nabla^{
\delta }H_{\alpha \beta \gamma } + \frac{1}{3} \nabla_{\delta 
}H_{\alpha \beta \gamma } \nabla^{\delta }H^{\alpha \beta 
\gamma }
\Big]\,.\labell{ST2}
\eeqa
One can write the above Lagrangian  as 
\beqa
\bL_e^{(1)}&=&\bL_M^{(1)} +\frac{2\lambda_0\alpha' }{\kappa^2} \nabla_\alpha(e^{-2\Phi}J^\alpha)\,,\labell{tot}
\eeqa
where $\bL_M^{(1)}$  is the  Meissner Lagrangian and the current in the total derivative term is 
\beqa
J^\alpha&=&8 R\nabla^{\alpha }\Phi + 16 \nabla^{\alpha }\Phi 
\nabla_{\beta }\nabla^{\beta }\Phi + 2 H^{\alpha \gamma 
\delta } H_{\beta \gamma \delta } \nabla^{\beta }\Phi - 16 
R^{\alpha }{}_{\beta } \nabla^{\beta }\Phi \nn\\&&- 16 
\nabla^{\alpha }\Phi \nabla_{\beta }\Phi \nabla^{\beta }\Phi 
- 16 \nabla_{\beta }\nabla^{\alpha }\Phi \nabla^{\beta }\Phi 
+ H^{\beta \gamma \delta } \nabla_{\delta }H^{\alpha 
}{}_{\beta \gamma } -  H^{\alpha \beta \gamma } \nabla_{\delta 
}H_{\beta \gamma }{}^{\delta }\,.
\eeqa
Note that neither the Lagrangian in the Meissner scheme nor the total derivative terms are individually invariant under T-duality. However, their combinations are invariant. Additionally, it is worth noting that while the Lagrangians \reef{fs3} and \reef{mie} are equivalent up to field redefinitions and total derivative terms, it can be demonstrated that the Lagrangians \reef{ST1} and \reef{ST2} are equivalent solely up to field redefinitions.

\subsection{Odd-parity coupling}

The T-duality of the odd-parity coupling in Equation \reef{CS} has been investigated in \cite{Garousi:2023pah} using the standard leading-order bulk and boundary actions, which are given as
\beqa
\bS^{(0)}+\partial\!\!\bS^{(0)}\!=\!-\frac{2}{\kappa^2}&\left[\int d^{D}x \sqrt{-G} e^{-2\Phi}\left( R\! +\! 4\nabla_{\mu}\Phi \nabla^{\mu}\Phi\!-\!\frac{1}{12}H^2\right) \!+\! 2\int d^{D-1}\sigma\sqrt{|g|} e^{-2\Phi}K\right],\labell{loa}
\eeqa
The last term above represents the Gibbons-Hawking boundary term \cite{Gibbons:1976ue}. In this case, the $O(1,1,\mathbb{Z})$ symmetry of the bulk coupling in Equation \reef{CS} necessitates  the following $\alpha'$-corrections of the Buscher rules\footnote{Note that the sign of the Taylor expansion of the leading-order action in \cite{Garousi:2023pah} is opposite to the sign in Equation \reef{TS11}. Moreover, the normalization of the Chern-Simons term in the field strength $\tilde H$ in Equation \reef{replace} is 3/2 compared to the one considered in \cite{Garousi:2023pah}.}:
\beqa
  \Delta \bar{g}_{ab}&=&0\,, \nn\\
  \Delta\bphi&=&0\,, \nn\\
\Delta\vp&=&\frac{1}{4}V^{ab}W_{ab}\,,\nn\\
\Delta g_{a}&=&-\frac{1}{16}\Big(2e^{\vp/2}\nabla^b\vp V_{ab}+2 e^{\vp/2}\bomega_{abc}V^{bc}-e^{-\vp/2}\bH_{abc} W^{bc} \Big)\,,\nn\\
\Delta b_{a}&=&-\frac{1}{16}\Big(2e^{-\vp/2}\nabla^b\vp W_{ab}-2 e^{-\vp/2}\bomega_{abc}W^{bc}+e^{\vp/2}\bH_{abc} V^{bc} \Big)\,,\nn\\
\tilde{B}_{ab}&=&0\,.\labell{dbH21}
\eeqa
as well as the presence of certain total derivative terms at order $\alpha'$ in the base space. These terms are canceled out by the transformation of the boundary term in Equation \reef{loa} under the aforementioned generalized Buscher rules.
However, in this section, we will be employing the leading-order bulk Lagrangian described by Equation \reef{S0b}. The difference between the reduction of the bulk Lagrangian in Equation \reef{loa} and the Lagrangian in Equation \reef{S0b} lies in the presence of the following total derivative term in the base space:
\beqa
 -\frac{2 }{\kappa^2}\sqrt{-\bg} \,   \nabla_a\Big[-e^{-2\bphi} \nabla^a\vp\Big]\,.
\eeqa
Using Stokes' theorem in the corresponding action, it produces the following boundary term in the base space:
\beqa
 \frac{2 }{\kappa^2} \,e^{-2\bphi}\sqrt{-\bg} \,   n_a\nabla^a\vp\,.
\eeqa
This term is precisely canceled out by the reduction of the boundary Lagrangian in Equation \reef{loa}. Therefore, if one utilizes the reduction of the leading-order bulk Lagrangian in Equation \reef{S0b}, the calculations of the T-duality invariance of the coupling in Equation \reef{CS} yield no residual total derivative term at order $\alpha'$ at all. That is, the coupling in Equation \reef{CS} is invariant under T-duality with the same corrections to the Buscher rules as in Equation \reef{dbH21}.
Therefore, the circular reduction of the odd-parity Lagrangian in Equation \reef{CS} remains invariant under the $O(1,1,\mathbb{Z})$ transformations with no anomalous term, as for the even-parity Lagrangian in Equation \reef{ST1} or Equation \reef{ST2}.

\section{Discussion}

In this paper, we demonstrate that by imposing the $O(1,1,\mathbb{Z})$ symmetry on the circular reduction of the most general covariant Lagrangian at order $\alpha'$, both the Lagrangian itself and the $\alpha'$-corrections to the Buscher rules can be determined. The fixing process involves one unambiguous parameter and 14 arbitrary parameters. The selection of these additional arbitrary parameters corresponds to different Lagrangian formulations in various schemes. Specifically, we have successfully identified T-duality invariant Lagrangians that correspond to the Metsaev-Tseytlin and Meissner Lagrangians. The difference between these Lagrangians and the T-duality invariant Lagrangians arises solely from specific total derivative terms. We anticipate that this observation holds true for effective Lagrangians at higher orders of $\alpha'$ as well.

In practice, dealing with higher orders of $\alpha'$ presents challenges when working with covariant couplings, where only the redundancy resulting from the Bianchi identities is removed. The sheer number of couplings involved makes it difficult to handle. As a solution, field redefinitions can be employed within the Lagrangian to eliminate the redundancy resulting from these field redefinitions as well. This effectively reduces the number of independent couplings. For example, at order $\alpha'$, instead of the 31 couplings considered in this paper, there are only 17 independent couplings. In this reduced scenario, T-duality can determine all parameters except for one unambiguous parameter and several arbitrary parameters, which can be set to zero to determine the couplings within a specific minimal scheme. Furthermore, after fixing the couplings in a specific minimal scheme, field redefinitions can still be utilized to express the effective Lagrangian in alternative schemes. Moreover, field redefinitions can be employed along with the removal of total derivative terms to express the effective action in particular schemes.

The T-duality constraint has been utilized in \cite{Garousi:2019mca, Garousi:2023kxw, Garousi:2020gio} to determine the effective actions up to order $\alpha'^3$. In these calculations, one starts with the independent couplings where redundancy arising from the Bianchi identities, field redefinitions, and total derivative terms has been eliminated. As a result, the number of independent couplings is reduced compared to the effective Lagrangians where the total derivative terms are not removed. However, when applying T-duality to the effective actions, it becomes necessary to include all possible total derivative terms in the base space. Since the massless fields in the base space are more numerous than those in the original spacetime, the sheer number of total derivative terms in the base space becomes overwhelming to consider. Therefore, we anticipate that at higher orders of $\alpha'$, the T-duality calculations involved in determining the effective Lagrangian will be significantly less complicated compared to those required for finding the effective actions.

We have observed that the $O(1,1,\mathbb{Z})$ symmetry can generate T-duality invariant Lagrangians and their corresponding generalized Buscher rules in various schemes. One can fix the scheme by specifying a particular form for the generalized Buscher rules. However, these specifications are subject to a constraint that requires the unambiguous parameter to remain non-zero. This logic can be employed to reduce the number of covariant couplings, as exemplified in equation \reef{all}, and the generalized Buscher rules, as seen in equation \reef{tH}, before imposing the $O(1,1,\mathbb{Z})$ constraint.
To decrease the number of couplings in the original Lagrangian at order $\alpha'$, one can first eliminate each term that involves three or four derivatives and then impose the Bianchi identities. This procedure leads to the identification of 26 independent couplings. The eliminated terms can indeed be transformed into other terms using integration by parts, which does not affect the unambiguous couplings in the Lagrangian. Consequently, the resulting Lagrangian and its circular reduction will only contain terms with up to two derivatives. Therefore, the generalized Buscher rules should not produce terms with more than two derivatives. By considering the fact that the second derivatives of the deformations $\Delta\bar{\phi}$ and $\Delta\bar{g}_{ab}$ appear in equation \reef{delS}, we find that these deformations generate terms with more than two derivatives. Hence, they should be set to zero in the corresponding generalized Buscher rules. Moreover, since the first derivative of all other deformations appears in equation \reef{delS}, they should involve only terms with first derivatives.
By applying this specific scheme, we find that the $O(1,1,\mathbb{Z})$ constraint fixes the generalized Buscher rules to be \reef{dbH1}, and the corresponding Lagrangian includes the unambiguous parameter $a_1$ and two other arbitrary parameters. By choosing specific values for these parameters, such that the pure gravity couplings become the Gauss-Bonnet couplings, we obtain
\beqa
\bL_e^{(1)}&=& -\frac{2\alpha' a_1}{\kappa^2} e^{-2\Phi}\sqrt{-G}\,\Big[H_{\alpha }{}^{\delta \epsilon } H^{\alpha \beta \gamma } 
H_{\beta \delta }{}^{\varepsilon } H_{\gamma \epsilon 
\varepsilon } - 3 H_{\alpha \beta }{}^{\delta } H^{\alpha \beta 
\gamma } H_{\gamma }{}^{\epsilon \varepsilon } H_{\delta 
\epsilon \varepsilon } \nn\\&&+ \frac{1}{6} H_{\alpha \beta \gamma } 
H^{\alpha \beta \gamma } H_{\delta \epsilon \varepsilon } 
H^{\delta \epsilon \varepsilon } + 48 H_{\alpha }{}^{\gamma 
\delta } H_{\beta \gamma \delta } R^{\alpha \beta } - 
96 R_{\alpha \beta } R^{\alpha \beta } - 4 
H_{\alpha \beta \gamma } H^{\alpha \beta \gamma } R + 
24 R^2 \nn\\&&+ 24 R_{\alpha  \beta \gamma \delta } 
R^{\alpha \beta \gamma \delta } -144 H_{\alpha 
}{}^{\delta \epsilon } H^{\alpha \beta \gamma } 
R_{\beta  \gamma \delta\epsilon } - 16 H_{\beta \gamma 
\delta } H^{\beta \gamma \delta } \nabla_{\alpha 
}\nabla^{\alpha }\Phi + 192 R \nabla_{\alpha }\nabla^{
\alpha }\Phi \nn\\&&+ 16 H_{\beta \gamma \delta } H^{\beta \gamma 
\delta } \nabla_{\alpha }\Phi \nabla^{\alpha }\Phi - 192 
R \nabla_{\alpha }\Phi \nabla^{\alpha }\Phi - 384 
R^{\alpha \beta } \nabla_{\beta }\nabla_{\alpha }\Phi 
+ 384 \nabla_{\alpha }\nabla^{\alpha }\Phi \nabla_{\beta 
}\nabla^{\beta }\Phi\nn\\&& - 768 \nabla_{\alpha }\Phi 
\nabla^{\alpha }\Phi \nabla_{\beta }\nabla^{\beta }\Phi - 96 
H_{\alpha }{}^{\gamma \delta } H_{\beta \gamma \delta } 
\nabla^{\alpha }\Phi \nabla^{\beta }\Phi + 384 \nabla_{\alpha 
}\Phi \nabla^{\alpha }\Phi \nabla_{\beta }\Phi \nabla^{\beta 
}\Phi\nn\\&& + 96 H_{\alpha }{}^{\gamma \delta } H_{\beta \gamma 
\delta } \nabla^{\beta }\nabla^{\alpha }\Phi - 384 
\nabla_{\beta }\nabla_{\alpha }\Phi \nabla^{\beta 
}\nabla^{\alpha }\Phi - 24 \nabla_{\alpha }H^{\alpha \beta 
\gamma } \nabla_{\delta }H_{\beta \gamma }{}^{\delta }\nn\\&& + 96 H_{
\alpha }{}^{\beta \gamma } \nabla^{\alpha }\Phi 
\nabla_{\delta }H_{\beta \gamma }{}^{\delta } + 8 
\nabla_{\delta }H_{\alpha \beta \gamma } \nabla^{\delta 
}H^{\alpha \beta \gamma }\Big]\,.\labell{sf1}
\eeqa
The Lagrangian \reef{ST2} is equivalent to this Lagrangian, up to the utilization of the Bianchi identities.

We have seen that the $O(1,1,\MZ)$ constraint can be used to find the T-duality invariant bulk Lagrangian up to one unambiguous parameter. One may expect that this constraint can also fix the covariant boundary Lagrangian. It has been shown in \cite{Garousi:2019xlf} that the following boundary couplings at the leading order of $\alpha'$ are invariant under the Buscher rules:
\beqa
\partial\bL^{(0)}&=&  -\frac{2a_5}{\kappa^2} e^{-2\Phi}\sqrt{|g|}\,  \left(-\frac{1}{2} K+n^{\mu}\nabla_{\mu}\Phi \right)\,,\labell{bbaction}
\eeqa
where $K$ represents the extrinsic curvature of the boundary. The sum of the bulk Lagrangian \reef{S0b} and the aforementioned boundary Lagrangian, with $a_5=-4a_1$, yields the standard Lagrangians in \reef{loa} after applying Stokes' theorem.

We have observed that by imposing the requirement for the most general covariant and gauge-invariant couplings in the effective Lagrangian at order $\alpha'$ to be compatible with T-duality, it is possible to determine the corrections to the Buscher rules and the effective Lagrangian up to one unambiguous parameter and 14 arbitrary parameters.
However, if the covariance of the couplings is not enforced, alternative couplings may exist that still maintain consistency with the aforementioned T-duality. In the context of the heterotic theory, the Lorentz-Chern-Simons form has been supersymmetrized, leading to the discovery of the following couplings in \cite{Bergshoeff:1988nn,Bergshoeff:1989de}:
\beqa
\bL_{BR}= -\frac{2}{\kappa^2} e^{-2\Phi}\sqrt{-G}\, \Big[ R -4\nabla_{\mu}\Phi \nabla^{\mu}\Phi-\frac{1}{12} \hat H^2+4\nabla_\mu\nabla^\mu\Phi+\frac{\alpha' }{8}R_{\mu\nu ij}(\Omega)R^{\mu\nu ij}(\Omega)\Big],\labell{BR}
\eeqa
where $\Omega_{\mu}{}^{ij}=\omega_{\mu}{}^{ij}-\frac{1}{2}\hat H_{\mu}{}^{ij}(\Omega)$, and $
R_{\mu\nu}{}^{ij}(\Omega)=\prt_\mu\Omega_\nu{}^{ij}-\prt_\nu\Omega_\mu{}^{ij}+\Omega_{\mu}{}^{ik}\Omega_{\nu k}{}^j-\Omega_{\nu}{}^{ik}\Omega_{\mu k}{}^j\,.$
This curvature can be expressed in terms of the standard curvature $R_{\mu\nu}{}^{ij}(\omega)$, the $\hat H^2$-terms, and a non-covariant but locally Lorentz-covariant derivative of $\hat H$. The presence of $\hat H(\Omega)$ in the Lagrangian indicates that it has couplings at all orders of $\alpha'$. It has been demonstrated in \cite{Bergshoeff:1988nn,Bergshoeff:1989de} that the aforementioned Lagrangian, at order $\alpha'$, is invariant under supersymmetry transformations. Additionally, it has been observed that these couplings also remain invariant under T-duality \cite{Bergshoeff:1995cg}.
Furthermore, it has been shown in \cite{Chemissany:2007he} that the aforementioned Lagrangian at order $\alpha'$ and the Metsaev-Tseytlin Lagrangian \reef{fs3} are equivalent up to certain total derivative terms and non-covariant field redefinitions. We have explicitly demonstrated the invariance of the above couplings at order $\alpha'$ under the Buscher rules in the reduction scheme \reef{reduc}. The corrections to the Buscher rules have been found to exclude any 2-form $\tilde B$, as well as corrections to the base space metric and dilaton. For a flat base space, the corrections are given by:
\beqa
\Delta\vp&=&\frac{1}{8}\Big(2\prt_a\vp\prt^a\vp+e^\vp V^2+e^{-\vp}W^2+2V^{ab}W_{ab}\Big)\,, \nn\\
\Delta g_{a}&=&\frac{1}{8}\Big(\frac{1}{2}e^{-\vp/2}\bH_{abc}W^{bc}-e^{\vp/2}V_{ab}\prt^b\vp +\frac{1}{2}e^{\vp/2}\bH_{abc} V^{bc}+e^{-\vp/2}W_{ab} \prt^b\vp \Big)\,,\nn\\
\Delta b_{a}&=&- \Delta g_{a}\,.\labell{dBR}
\eeqa
We have examined the invariance of the couplings at order $\alpha'^2$ under T-duality and have not observed their preservation. This observation may  suggest the potential existence of additional couplings at this order in the aforementioned Lagrangian. Uncovering such couplings would be intriguing, and one approach to achieve this is by imposing the constraint that the circular reduction of the Lagrangian remains invariant under $O(1,1,\mathbb{Z})$ transformations. 
The newly introduced couplings should also be consistent with supersymmetric transformations at the order of $\alpha'^2$. According to supersymmetry, the new terms should not include Riemann cubed terms \cite{Bergshoeff:1988nn}.

\end{document}